\documentclass[reprint,superscriptaddress, amsmath,amssymb,aps,longbibliography]{revtex4-1}

 \usepackage[bottom]{footmisc}

\usepackage[T1]{fontenc}
\usepackage[latin9]{inputenc}
\usepackage{mathrsfs}
\usepackage{bm}
\usepackage{amsmath}
\usepackage{amsthm}
\usepackage{amssymb}
\usepackage{stmaryrd}
\usepackage{mathtools}
\usepackage{dsfont}
\newcommand{\ud}{\mathrm{d}}

\newcommand{\bt}{\boldsymbol{\theta}}

\newcommand{\Wp}{{\mathcal{P}}}

\edef\ordinarycolon{\mathchar\the\mathcode`: }
\edef\ordinaryequals{\mathchar\the\mathcode`= }

\usepackage{breqn}
\catcode`^=7

\AtBeginDocument{%
  \catcode`^=12
}

\makeatletter

\allowdisplaybreaks

\let\cat@comma@active\@empty

\usepackage[capitalize]{cleveref}

\usepackage{color}
\RequirePackage[dvipsnames,usenames]{xcolor}

\newif\ifnotes
\notestrue

 \newcommand{\jan}{\color{black}}

\makeatother

\newcommand{\ba}{\begin{eqnarray}}
\newcommand{\ea}{\end{eqnarray}}

\newcommand{\eq}[1]{\begin{align}#1\end{align}}

\begin{document}

\preprint{}

\title{Thermodynamics of nonequilibrium systems with uncertain parameters}

\author{Jan Korbel}
 \affiliation{Section for Science of Complex Systems, CeMSIIS, Medical University of Vienna, Spitalgasse 23, 1090 Vienna, Austria}
\email{jan.korbel@meduniwien.ac.at}
\affiliation{Science Hub Vienna, Josefst\"{a}dter Strasse 39, 1080 Vienna, Austria}
\author{David H. Wolpert}%
\affiliation{Santa Fe Institute, Santa Fe, NM, United States of America}
\email{david.h.wolpert@gmail.com}
\homepage{ http://davidwolpert.weebly.com}
\affiliation{Science Hub Vienna, Josefst\"{a}dter Strasse 39, 1080 Vienna, Austria}
\affiliation{ Arizona State University, Tempe, AZ, United States of America}






\date{\today}

\begin{abstract}
{ In the real world, one almost never knows the parameters of a thermodynamic process to infinite precision.
Reflecting this, here we investigate how to extend stochastic thermodynamics to systems with uncertain parameters, including
uncertain number of heat baths / particle reservoirs, uncertainty in the precise values of
temperatures / chemical potentials of those reservoirs, uncertainty in the energy spectrum, uncertainty in the
control protocol, etc.
We formalize such uncertainty
with an (arbitrary) probability measure over all transition rate matrices satisfying local detailed balance.
This lets us define the effective thermodynamic quantities by averaging over all LDB-obeying  rate matrices.
We show that the resultant effective entropy violates the second law of thermodynamics.
In contrast to the effective entropy though, the expected stochastic entropy,
defined as the ensemble average of the effective trajectory-level entropy, satisfies the second law.
We then and explicitly calculate the second-order
correction to the second law for the case of one heat bath with uncertain temperature.
We also derive the detailed fluctuation theorem for expected effective trajectory entropy production
for this case, and derive a lower bound
for the associated expected work. Next, to ground these formal considerations with experimentally testable bounds on allowed
energetics, we derive a bound on the maximal work that can be extracted from systems with arbitrarily uncertain temperature.
We end by extending previous work on ``thermodynamic value of information'', to allow for uncertainty in
the time-evolution of the rate matrix.}

\end{abstract}

\maketitle
\section{Introduction}
The microscopic laws of classical and quantum physics provide us a parameterized set of equations
that specify the evolution of a closed system starting from a specific state. To use those equations we
need to know that specific state, we need to be sure the system is closed, and we need to know the
values of the parameters in the equations.~\cite{barato2015}

Unfortunately, in very many real-world scenarios, we are uncertain about the precise state of the system, and
very often the system is open rather than closed, subject to
uncertain interactions with the external environment.
Statistical physics captures these two types of uncertainty by building on the microscopic laws of physics in two ways.
First, to capture uncertainty about the state of the system,
in statistical physics we replace exact specification of the system's state with a probability distribution
over states. Second, to capture uncertain interactions between the system and the external environment, we add randomness to the
dynamics, in a precisely parameterized form~\footnote{We note though there is a substantial
literature in which adopts an ``inclusive Hamiltonian'' framework, in which the external environment is finite, and
the joint dynamics of the system-environment is explicitly
modeled. This has been explored either with explicit Hamiltonian dynamics, with only randomness in the initial
state~\cite{kawai_dissipation:_2007,esposito2010entropy,ptaszynski2018first} or at least approximate Hamiltonian
dynamics~\cite{seifert2016first,strasberg2017stochastic,talkner2020HMF,talkner2020comment,strasberg2020measurability}.}.

In particular, in classical stochastic thermodynamics~\cite{seifert2012stochastic,van2015ensemble},
we model the system as a probability distribution evolving under a continuous-time Markov chain (CTMC),
with a precisely specified rate matrix. Typically we require that the CTMC obeys local detailed balance (LDB). This
means that the rate matrix of the CTMC has to obey certain restrictions which are parameterized by the energy
spectrum of the system, the number of thermodynamic reservoirs in the external environment perturbing
the system's dynamics, and the temperatures and chemical potentials of those reservoirs. Often we also
allow the rate matrix of the CTMC to change in time, subject to those restrictions, according to a precisely defined ``protocol''.

However, in addition to uncertainty about the state of the system, and uncertainty about interactions
with the external environment, there is an additional unavoidable type of uncertainty in all real-world systems: uncertainty
about the parameters in the equations governing the dynamics. In the context of stochastic thermodynamics, this
means that even if we impose LDB, we will \textit{never} know the reservoir temperatures and
chemical potentials to infinite precision (often even being unsure about the number of such reservoirs),
we will never know the energy spectrum to infinite precision, and more generally we will never know the rate matrix
and its time-dependence to infinite precision.

In this paper we consider how stochastic thermodynamics (and nonequilibrium statistical physics more
generally) needs to be modified to capture this third type of uncertainty, in
addition to the two types of uncertainty it already captures. We emphasize that this third type of uncertainty,
concerning the parameters of the thermodynamic process, is unavoidable in essentially
all real-world scenarios. Nonetheless, almost nothing is known about its thermodynamic consequences~\footnote{We
note though that it is
already known that if we do not account for all thermodynamic reservoirs, we invariably undercount the total
of entropy production in a process~\cite{esposito2010three}.}.

To clarify our precise focus in this paper, we note that there has been some preliminary research
on how to modify stochastic thermodynamics in the case that we, the experimentalist, are not able to
view all state transitions in the system as it evolves~\cite{bisker2017hierarchical,shiraishi_ito_sagawa_thermo_of_time_separation.2015}.
In contrast to our focus, this uncertainty concerns what is observed as the system evolves, whereas
the uncertainty we consider involves the parameters governing
that evolution. Similarly, some models consider either spatial \cite{matsuo2000stochastic} or temporal \cite{fiore2019entropy} variation of temperature and other parameters, but they assume that this evolution is known.

{ Probably the closest research to what we consider in this paper
 is sometimes called \emph{superstatistics}. It has long been known that an average over Gibbs distributions cannot be written
as some single Gibbs distribution (Thm.\,1 in~\cite{wolpert1993use}). This means that even equilibrium statistical physics
must be modified when there is uncertainty in the temperature of a system.
The analysis of these modifications was begun by Beck and Cohen \cite{beck2003superstaitstic},
who developed an effective theory for thermodynamics with temperature fluctuations. They considered a system coupled to a bath,
which is in a \emph{local equilibrium} under the slow evolution of the temperature of the bath.
The main assumption they exploit is \emph{scale-separation}: while for short time scales, the equilibrium distribution is a canonical distribution with inverse temperature $\beta$, the long-scale behavior is determined by the \emph{superposition} of canonical distributions with some distribution of temperatures $f(\beta)$.
The resulting \emph{superstatistical} distribution $p(E) = \int \mathrm{d} \beta f(\beta) \exp(-\beta E)$ was later
identified with the distribution corresponding to generalized entropic functionals \cite{hanel2011generalized,garcia2011superstatistics}.
Later interpretations of superstatistics were not based on the notion of local equilibria but rather on the
Bayesian approach to systems with uncertain temperature \cite{sattin2006bayesian,davis2018temperature}. These are
conceptually closer to the focus of this paper.}


The rest of the paper is organized as follows: in the next section, we introduce necessary notation and briefly recall the main results of stochastic thermodynamics. In \cref{sec:3}, we present the general framework of expected thermodynamics for systems with uncertain transition rates. {First we note that the expected probability distribution is not Markovian.
We then show that the expected entropy does not satisfy the second law.
On the other hand, the \emph{expected stochastic entropy}, defined as the ensemble average of the
expected trajectory entropy, does satisfy the second law.}

For the remainder of the paper, we focus on the simple case where we know for
certain that there is a single (heat) reservoir, but are uncertain about its temperature, which is constant during the evolution of the system.
{ The uncertainty in temperature implies uncertainty in the transition rate matrix, since we mean assume that
the actual rate matrix, whatever it is, satisfies LDB.}
In  \cref{sec:4}, we consider the special case where the temperature distribution is sharply peaked.
{ We show that the expected entropy flow can be smaller than expected inverse temperature times expected heat and explicitly calculate the correction to the second law. Furthermore, we derive the detailed fluctuation theorem for the expected trajectory entropy production. Finally, for the case of equilibrium initial and final distributions, we show that the expected work is not lower-bounded by the expected difference in the equilibrium Helmholtz free energy, but derive a correction to the lower bound.}

In the final two sections we consider an issue foundational to stochastic thermodynamics: How much work
can be extracted from a system during a process that takes it from its initial distribution to a given,
target distribution. First, in \cref{sec:first_minimal_EP_section}, we consider this issue when we are uncertain both about the
initial distribution and about the temperature of the system as it evolves. We focus on how that uncertainty changes the results of the
standard analysis of this issue, in which we suppose a \{quench; equilibrate; semistatically-evolve\} process is applied
to the system immediately after the initial distribution is generated~\cite{parrondo2015thermodynamics,hasegawa2010generalization}.
Then in \cref{sec:second_minimal_EP_section}, we simplify this scenario. First, we suppose that
the temperature during the dynamics is perfectly known, so that the only uncertainty is in the initial distribution. However,
we also suppose that there is a (perfectly known) delay between when the initial distribution is generated, $t_i$, and
the time $\tau$ when
the  \{quench; equilibrate; semistatically-evolve\} process can begin, during which time the system evolves according
to a (perfectly known) rate matrix. Specifically, we analyze how the minimal EP changes as the length of the delay changes,
which causes the distributional uncertainty at the beginning of the quench process to change. We then present preliminary
results of how this analysis changes when there is uncertainty about the rate matrix taking the initial distribution to
the time $\tau$ distribution, in addition to (and possibly statistically coupled with) the uncertainty about
the initial distribution.

We end with a discussion section in which we describe just a few of the myriad directions for future work.


\section{Terminology and notation}
\label{sec:2}

In conventional stochastic thermodynamics, with no uncertainty about the thermodynamic parameters,
we consider a system coupled to $N$ heat and/or particle reservoirs, with temperatures $\beta_1,\dots,\beta_N$ and chemical potentials $\mu_1,\dots,\mu_N$. As shorthand write $\theta^\nu =(\beta_\nu,\mu_\nu)$ and $\bt = (\theta^1,\dots,\theta^N)$.
We write the rate matrix for going from state $x'$ to state $x$ as $K_{x,x'}$, with the time index implicit.
So the master equation is $\dot{p}_x(t) = \sum_{x'} K_{x,x'} p_{x'}(t)$. As usual, we assume that the
rate matrix can be decomposed as $K_{x,x'} = \sum_{\nu=1}^{N} K_{x,x'}^{\nu}$ where
the separate matrices $K_{x,x'}^{\nu}$ each satisfy LDB.

The internal energy is written as $U = \sum_x p_x \epsilon_x$, and the first law of thermodynamics is written in
the usual way, as $\dot{U} = \dot{W} + \dot{Q} + \dot{W}_{chem}$, where $\dot{W}(t) = \sum_x \dot{\epsilon}_x(t) p_{x}(t)$.
The heat flow $\dot{Q}$
is decomposed into heat flows to particular reservoirs:
\eq{
\dot{Q}(t) &= \sum_{\nu} \dot{Q}_\nu(t) \\
	&= \sum_{\nu} \sum_x (\epsilon_x(t)-\mu_\nu n_x) K_{x,x'}^{\nu} p_x(t)
}
where $n_x$ is the number of particles in the system, as specified by the state $x$. (For simplicity, we ignore
the possibility of more than one distinguishable type of particle.)
The chemical work is then given by $\dot{W}_{chem} = \sum_\nu \sum_x \dot{p}_x(t) \mu_\nu n_x$.
The Shannon entropy is
written as $S = - \sum_x p_x \ln p_x$. The second law of thermodynamics can be expressed as $\dot{S} = \dot{S}_i + \dot{S}_e$ where $ \dot{S}_i \geq 0$ and $\dot{S}_e =  \sum_\nu  \beta_\nu \dot{Q}^{\nu}$.

Throughout this paper, for simplicity we assume that any time-extended thermodynamic process we refer to occurs in some
interval $[t_i, t_f]$ where we have no uncertainty about those starting and ending times.
We write a generic stochastic trajectory across that interval as $x(t)$. So the
\textit{stochastic energy} at time $\tau$ is $\epsilon(t) = \epsilon_{x(t)}$,
and the first law of thermodynamics on the trajectory level for any
time $t$ is $\frac{\ud}{\ud t} e(t) = \dot{w}(t) + \dot{q}(t) + \dot{w}_{chem}(t)$,
where
\eq{
\dot{w}(t) &= \sum_x \delta_{x,x(t)} \dot{\epsilon}_{x(t)}, \\
\label{eq:4}
\dot{q}(t) &= \sum_\nu \dot{q}^\nu(t)= \sum_{\nu}  \dot{\delta}_{x,x(t)} (\epsilon_{x(t)}-\mu_\nu n_{x(t)}), \\
\dot{w}_{chem} &= \sum_\nu  \dot{\delta}_{x,x(t)} \mu_\nu n_{x(t)}
}
are called the \textit{stochastic work, stochastic heat}, and \textit{stochastic chemical work}, respectively.

As usual, the \textit{stochastic entropy} along a trajectory is defined as $s_{x(\tau)} = - \ln p_{x(\tau)}$.
Stochastic entropy production is then
\eq{
\dot{s}_i = \dot{s} -  \dot{s}_e
\label{eq:stoch_entr_fixed_rate_matrix}
}
where  due to LDB, $\dot{s}_e = \sum_\nu  \beta_\nu \dot{q}^\nu$.
Due to LDB, stochastic entropy production fulfills the \emph{detailed fluctuation theorem},
$P(\Delta s_i)/\tilde{P}(-\Delta s_i) = e^{\Delta s_i}$, where $\tilde{P}$ denotes the probability under a time-reversed protocol. By averaging these trajectory-level quantities over all trajectories, we recover the ensemble-level versions. For more details, see the Appendix.

\section{Effective thermodynamics of the system with uncertain transition rates}
\label{sec:3}

As described in the introduction, we are interested in how the conventional laws of stochastic thermodynamics concerning
the evolution of a system change when there is uncertainty about the parameters of that evolution, but we
assume LDB holds, whatever those parameters are.
We start with comments on the general case where the transition rates of the systems are not known with
infinite precision. Such uncertainty must
arise whenever we do not know exact number of heat reservoirs, their temperatures/and or chemical potentials, but
can even arise when we \textit{do} know those quantities exactly, and even when we impose LDB.
(For simplicity, we not consider uncertainty in the energy spectrum or how it varies in time. \footnote{{\jan Uncertainty in the energy spectrum and / or the time-dependence of the rate matrix can also be included in the analysis, at the cost of more complex notation.}})
However, note even if we knew those quantities to infinite precision, LDB does not uniquely
fix the rate matrix $K$, and so there can still be uncertainty
concerning $K$.

In the Appendix, 
we review why the general form of a transition matrix satisfying LDB can be written as
\eq{
&K(N;\bt,\mathbf{R}) \;=\; \sum_\nu K^\nu(\theta^\nu,R^\nu) \nonumber\\
&\;\; = \sum_{\nu=1}^N \left[R^\nu \, (\Pi^\nu(\theta^\nu))^{-1}
\mathrm{diag} \left(R^\nu \, (\Pi^\nu(\theta^\nu))^{-1} \cdot \textbf{1}\right)\right]
\label{eq:LDB}
}
where each $R^\nu$ is an arbitrary symmetric matrix and the associated $\Pi^\nu$ is a diagonal matrix whose diagonal elements
are the entires in the equilibrium distribution of the reservoir $\nu$. 

We now consider a probability measure $\mathrm{d} F_K$ over the rate matrices obeying
\cref{eq:LDB}, which quantifies our uncertainty about the temperatures / chemical potential of the reservoirs, number of reservoirs, etc.
We suppose that before a system starts to evolve, its rate matrix is generated by sampling $\ud F_K$, and that
the system then evolves with that rate matrix, without ever again sampling $\ud F_K$. However, the experimentalist
is still uncertain what that randomly generated rate matrix is as they observe the system evolving.
We emphasize that this scenario differs from one where the system couples / decouples from the reservoirs
in a random process, or where the reservoirs undergo temperature fluctuations during its evolution, as for example as
considered in superstatistics~\cite{beck2003superstaitstic}.

We write the associated \emph{effective} value of a (perhaps implicitly time-dependent) generic quantity $X^{K}$ as
\begin{eqnarray}
\overline{X} :&=& \int \ud F_{K} \, X^K\nonumber\\
&=& \sum_{N=1}^\infty \Wp_N \int \prod_{\nu=1}^N \mathrm{d} \mathbf{R}^\nu \mathrm{d} \boldsymbol{\theta}^\nu \,  f^N(\bt,\mathbf{R})\, X^{K(N;\bt,\mathbb{S})}
\end{eqnarray}
where $\Wp_N$ is the probability distribution over the number of reservoirs and $f^N(\bt,\mathbf{R})$ is the distribution over all
possible LDB-obeying rate matrices for a system coupled with $N$ reservoirs.
We use the term ``effective value'' to refer to expectations formed this way by
integrating over all possible values of $N, \boldsymbol{\theta}, \mathbf{R}$.
We instead use the term \emph{average} value to mean expectations formed by integrating
over all possible ensemble trajectories, and denote such a value generically as $\langle \cdot \rangle$.

We emphasize that there are other types of uncertainty concerning thermodynamic parameters,
reflecting imperfect knowledge of energy levels, chemical potentials, the control protocol driving the energy spectrum, etc.
For simplicity, in this paper we do not consider these other types of parameter uncertainty. However, it is
straight-forward to generalize some of the results below to capture these other types of uncertainty as well.


%
%
%
Using this notation, the {effective} distribution at any time $t$ is
\begin{equation}
\overline{p}_x(t) := \int \ud F_{K} p^{K}_x(t)\, .
\end{equation}
and the time evolution of $\bar{p}$ is given by
\begin{equation}
\dot{\overline{p}}_x(t) = \sum_{x'} \overline{J}_{x,x'} = \sum_{x'} \int \ud F_{K} K_{x,x'} p^{K}_{x'}(t)
\end{equation}

{ The corresponding transition matrix  at time $t$ can be formally defined as
\begin{equation}
\mathcal{K}_{x,x'}(p^K(t)) := \frac{\int \ud F_{K} K_{x,x'} p^{K}_{x'}(t)}{\int \ud F_{K} p^{K}_{x'}(t)}
\end{equation}
depends on all $p^K$.}
However, this matrix will change if the distributions $p^K_{x'}(t)$ change, i.e., is not only a function
of the state $x$. Accordingly, it results in non-Markovian dynamics~\cite{strasberg2019non}.
Note as well
that $\overline{K} = \int \ud F_{K}\, K$, the effective transition rate matrix, is not directly related to the
dynamics of the effective probability distribution.
In particular $\overline{p}_x(t) \neq p_x^{\overline{K}}(t)$ in general.


Next, write the effective internal energy as $\overline{U} = \sum_x \overline{p}_x \epsilon_x$.  The first law of thermodynamics in
terms of this effective internal energy
can be expressed as $\dot{\overline{U}} = \dot{\overline{W}} + \dot{\overline{Q}} + \dot{\overline{W}}_{chem}$. Similarly, we can introduce the effective entropy as $\bar{S} = - \sum_x \overline{p_x \ln p_x}$ and write down the second law of thermodynamics as
\begin{equation}
\dot{\bar{S}} = \dot{\bar{S}}_i + \dot{\bar{S}}_e
\end{equation}
 where
\eq{
\label{eq:12a}
\dot{\bar{S}}_i &= \int \ud F_K \dot{S}^K_i \geq 0 \\
\dot{\bar{S}}_e &=   \int \ud F_K \dot{S}^K_e \\
	&=  \overline{\sum_{\nu = 1}^N \beta^K_\nu \dot{Q}^{\nu,K}}
\label{eq:8_EF}
}
(Note that we average over $N$, and therefore over the number of terms in the sum in \cref{eq:8_EF}.)
$\dot{\bar{S}}_i$ is the effective entropy production rate, and $\dot{\bar{S}}_e$ is the effective entropy flow rate.

Note that the effective entropy production is non-negative, as in conventional
(no uncertainty) stochastic thermodynamics, since each $\dot{S}^K_i$ is non-negative. However,
there is no simple relation between the effective entropy flow rate and the effective heat flow rate, given by
\begin{equation}
\dot{\overline{Q}} = \overline{\sum_{\nu = 1}^N \dot{Q}^{\nu,K}}
\end{equation}
(This discrepancy between the two rates would exist even if we knew with
certainty that $N=1$; it reflects the fact that there is statistical
coupling between $\beta^K$ and $\dot{Q}^K$ when $\beta^K$ is a random variable.)
In particular, the derivative of the entropy is not necessarily lower-bounded by the heat flow rate, which
means that this ensemble version of the second law of thermodynamics is not as consequential as the standard version.

 We now turn our focus to trajectory-level thermodynamics. To begin, note that for a given trajectory $x(t)$
and given energy spectrum, the stochastic energy does not depend on the transition rate matrix.
Therefore stochastic work and
stochastic total heat (heat plus chemical work) are independent of the rate matrix.

Next, define the effective stochastic entropy at time $\tau$ as
\begin{equation}
\overline{s}(x(\tau)) = - \overline{\log \, p_{x(\tau)}}
\end{equation}
Note that the effective stochastic entropy is not the same as the stochastic entropy of the effective distribution,
$ - \log \bar{p}_x(t)$. If we
average the trajectory-level decomposition of entropy, \cref{eq:stoch_entr_fixed_rate_matrix},
over all possible transition rate matrices, we obtain
\begin{equation}\label{eq:sle1}
\dot{\bar{s}}(x(\tau)) = \dot{\bar{s}}_i(x(\tau)) + \dot{\bar{s}}_e(x(\tau))
\end{equation}
where
\eq{
\dot{\bar{s}}_e(x(\tau)) &= \int \ud F_K {\dot{s}}^K_e(x(\tau)) \nonumber\\
    &=\sum_{N=1}^\infty \Wp_N  \sum_{\nu=1}^N \overline{\beta_\nu (\epsilon_{x(\tau)} - \mu_\nu n_{x(\tau)})}\nonumber\\
     &=\sum_{N=1}^\infty \Wp_N  \sum_{\nu=1}^N \overline{\beta_\nu} (\epsilon_{x(\tau)} - \overline{\mu_\nu} n_{x(\tau)})\nonumber\\
	&= \sum_{N=1}^\infty \Wp_N  \sum_{\nu=1}^N \overline{\beta^\nu} \, \dot{\overline{q^\nu}}(x(\tau))
}
where in the last line we assume that $\mathrm{Cov}(\beta^\nu,\mu^\nu) = 0$, i.e., that our uncertainty about the
temperature is statistically independent of our uncertainty about the chemical potential.

By averaging \cref{eq:sle1} over all possible trajectories under the effective probability distribution $\overline{p}_{x(\tau)}$, we obtain
\begin{equation}\label{eq:sle}
\langle \dot{\bar{s}} \rangle_{\bar{p}} = \langle \dot{\bar{s}}_i \rangle_{\bar{p}} + \langle \dot{\bar{s}}_e \rangle_{\bar{p}}
\end{equation}
where we define $\langle f(x(\tau)) \rangle_{\bar{p}} := \sum_{x(\tau)} f(x(\tau)) \bar{p}_{x(\tau)}$. Note that
\begin{equation}
\dot{\bar{S}}_e = \overline{ \langle \dot{s}_e \rangle_{p^K}} \neq \langle \dot{\bar{s}}_e \rangle_{\bar{p}} = \sum_{N=1}^\infty \Wp_N  \sum_{\nu=1}^N \overline{\beta^\nu} \, \dot{\overline{Q^\nu}}
\end{equation}
{ As shorthand, write the effective stochastic entropy as
\begin{equation}
\mathcal{S} := \langle{\bar{s}} \rangle_{\bar{p}} = - \sum_x \overline{p}_x \, \cdot \, \overline{\ln p_x}
\end{equation}
With help of \cref{eq:sle} we can show that the effective stochastic entropy also satisfies the second law.
In addition, the effective stochastic entropy exceeds the stochastic entropy, $\overline{S}$:}
\begin{equation}\label{eq:2lt}
\dot{\mathcal{S}} \geq \sum_{N=1}^\infty \Wp_N  \sum_{\nu=1}^N \overline{\beta^\nu} \, \dot{\overline{Q^\nu}}\,.
\end{equation}




\section{System coupled to one reservoir with sharply peaked temperature distribution}
\label{sec:4}

In this section we consider the simple scenario where we are certain that
the system is coupled to a single heat reservoir, i.e., $\bt = \beta$
but are uncertain about the
temperature of that reservoir. Via LDB, this uncertainty in temperature causes uncertainty in the rate matrix,
and therefore uncertainty in the stochastic thermodynamics.
For simplicity, we only consider the effect of this specific type of uncertainty on expectation values.
We indicate such an effective value of a general thermodynamic variable $X(\beta)$ (possibly depending on the trajectory $x(t)$ defined in the time interval $[t_i, t_f]$) as
\begin{equation}
\bar{X} = \int \ud F_\beta X(\beta)
\end{equation}
%

To begin, we make a second-order expansion of $X$ in $\beta-\bar{\beta}$ to obtain
\begin{equation}
\bar{X} \approx X(\bar{\beta}) + \frac{\mathrm{Var} \beta}{2} \frac{\partial^2 X}{\partial \beta^2}\vert_{\bar{\beta}}
\label{eq:21}
\end{equation}
In particular,  the effective heat flow to second order is
\begin{equation}
\bar{Q} \approx Q(\bar{\beta}) + \frac{\mathrm{Var} \beta}{2} \frac{\partial^2 Q}{\partial \beta^2}\vert_{\bar{\beta}}
\label{eq:qb}
\end{equation}
Moreover, $\bar{S}_e$ can be expanded as
\begin{eqnarray}
\bar{S}_e &=& \overline{\beta Q} = \int \ud \beta \, p(\beta) \, \beta  Q(\beta) \nonumber\\
&\simeq& \int \ud \beta \,  p(\beta) \beta\left[Q(\bar{\beta}) + (\beta-\bar{\beta}) \frac{\partial Q}{\partial \beta}\vert_{\bar{\beta}} + \frac{(\beta-\bar{\beta})^2}{2} \frac{\partial^2 Q}{\partial \beta^2}\vert_{\bar{\beta}}\right]\nonumber\\
&=& \bar{\beta} Q(\bar{\beta}) + \mathrm{Var} \beta \frac{\partial Q}{\partial \beta}\vert_{\bar{\beta}} + \frac{\overline{\beta^3}-\overline{\beta}\, \overline{\beta^2}}{2} \frac{\partial^2 Q}{\partial \beta^2}\vert_{\bar{\beta}}
\end{eqnarray}
By plugging in for $Q(\bar{\beta})$ from \cref{eq:qb}, we obtain 
\begin{eqnarray}
\bar{S}_e = \bar{\beta} \bar{Q} + \mathrm{Var} \beta \frac{\partial Q}{\partial \beta}\vert_{\bar{\beta}} + \frac{\overline{\beta^3} - \bar{\beta}^3}{2} \frac{\partial^2 Q}{\partial \beta^2}\vert_{\bar{\beta}}\nonumber\\
= \bar{\beta} \bar{Q} - m_2(\beta) C_V + m_3(\beta) \left(C_V' + \frac{C_V''}{2\bar{\beta}}\right)
\end{eqnarray}
{ where $C_V = \frac{\partial Q}{\partial T}$ is the specific heat at $\bar{T} = \frac{1}{\bar{\beta}}$ and
\begin{equation}
m_k(\beta) = \frac{\overline{\beta^k}}{\bar{\beta}^k}-1
\end{equation}
is the \emph{relative $k$-th moment}. {It is straightforward to show from Jensen's inequality that $m_k(\beta) \geq 0$.}

Note that for many cases the effective entropy flow is smaller than effective inverse temperature times effective heat.
Consider, for example the case where $C_V = const >0$, i.e., positive specific heat independent of temperature.)
This can be interpreted as a {violation of the second law for \textit{effective thermodynamic quantities}}.
Formally, it arises because $\beta$ and $Q$ are statistically coupled, so
$\overline{\beta Q} \ne \overline{\beta} \; \overline{Q}$. It is important to
emphasize that this is not a \emph{physical violation} of the second law. Experimentally, it would arise if we repeatedly prepare the system
the same (imprecise) way, and interpret the resulting empirical average values of thermodynamic quantities
as an average under one, fixed distribution, being repeatedly sampled in each experiment, when in
fact the distribution itself varies from one experiment to the next.



We now use these preliminary results to investigate the trajectory thermodynamics when
the distribution over temperature is highly peaked. First, we note that
the effective change of entropy production $\bar{\Delta s}_i$ across a trajectory $x(\tau)$ starting at $t_0$ and ending at $t_f$  can be expressed as
\begin{equation}
\Delta \bar{s}_i = \overline{\ln \frac{\mathcal{P}^\beta(x(\tau))}{\tilde{\mathcal{P}}^\beta(\tilde{x}(\tau))}}
\end{equation}
where $\mathcal{P}^\beta$ is the probability of observing trajectory $x(\tau)$ for inverse temperature $\beta$, $\tilde{x}(\tau) = x(t_f - \tau)$ is the time-reversed trajectory $\tilde{\mathcal{P}}$ is the probability for the time-reversed protocol $\tilde{\lambda}(x)$.
{ We assume that for each $\beta$, we are certain about the initial and final distribution.}
By using expansion \cref{eq:21} we can express the effective entropy production change as:

\begin{equation}\label{eq:26}
\Delta \bar{s}_i = \Delta s_i(\bar{\beta}) + \frac{\mathrm{Var} \beta}{2} \frac{\partial^2 \Delta s_i}{\partial \beta^2}\vert_{\bar{\beta}}\, .
\end{equation}
Furthermore, due to the fact that the trajectory entropy flow is a linear function of $\beta$, we can use the following identity:
\begin{equation}
\left. \frac{\partial^2\Delta s_i}{\partial \beta^2} \right\vert_{\beta=\bar{\beta}} = \left. \frac{\partial^2 \Delta s}{\partial \beta^2} \right\vert_{\beta=\bar{\beta}}\, .
\end{equation}

An interesting special case is where the initial and final distribution are equilibrium distributions, as for example in Crooks' theorem.
So the equilibrium distribution for each $\beta$ is given by the Boltzmann distribution with corresponding temperature.

In this special case,
\begin{eqnarray}
\frac{\partial^2 s}{\partial \beta^2}\left.  \right\vert_{\beta=\bar{\beta}} &=& \frac{\partial^2}{\partial \beta^2} \left(\beta \epsilon_x + \ln Z\right) \left.\right\vert_{\beta=\bar{\beta}} \nonumber\\
&=& \mathrm{Var}(U)_{\bar{\beta}} = \sum_x \pi^{\bar{\beta}}_x \epsilon^2_x - \left(\sum_x \pi^{\bar{\beta}} \epsilon^2_x\right)^2\, .
\end{eqnarray}
so we finally obtain
\begin{equation}\label{eq:si}
\Delta \bar{s}_i = \Delta  s_i(\bar{\beta}) + \frac{1}{2} \mathrm{Var}(\beta)\, \Delta \mathrm{Var}(U)_{\bar{\beta}}\, .
\end{equation}
Depending on $\Delta \mathrm{Var}(U)_{\bar{\beta}}$, the trajectory entropy production can be smaller or bigger than for the case of effective temperature.

Expressing $\Delta s_i(\bar{\beta})$ from \cref{eq:26} and plugging it into the DFT for the entropy production change with effective temperature $\ln \frac{\mathcal{P}^{\bar{\beta}}(x(\tau))}{\tilde{\mathcal{P}}^{\bar{\beta}}(\tilde{x}(\tau))} = \Delta s_i(\bar{\beta})$, it is possible to obtain the detailed fluctuation theorem for $\bar{s}_i$  as
\begin{widetext}
\begin{eqnarray}
P^{\bar{\beta}}(\Delta s_i) = \int \mathcal{D} x(\tau) \mathcal{P}^{\bar{\beta}}(x(\tau)) \delta\left(\Delta \bar{s}_i - \overline{\ln \frac{\mathcal{P}^\beta(x(\tau))}{\tilde{\mathcal{P}}^\beta(\tilde{x}(\tau))}}\right)\nonumber\\
= e^{\Delta s_i - \frac{1}{2} \mathrm{Var}(\beta)\, \Delta \mathrm{Var}(U)_{\bar{\beta}}}\int \mathcal{D} x(\tau) \tilde{\mathcal{P}}^{\bar{\beta}}(\tilde{x}(\tau)) \delta\left(-\Delta \bar{s}_i - \overline{\ln \frac{\tilde{\mathcal{P}}^\beta(\tilde{x}(\tau))}{\mathcal{P}^\beta(x(\tau))}}\right) = e^{\Delta s_i - \frac{1}{2} \mathrm{Var}(\beta)\, \Delta \mathrm{Var}(U)_{\bar{\beta}}} \tilde{P}^{\bar{\beta}}(-\Delta s_i)
\end{eqnarray}
\end{widetext}
so we obtain
\begin{equation}
\frac{P^{\bar{\beta}}(\Delta s_i)}{ \tilde{P}^{\bar{\beta}}(-\Delta s_i)} = e^{\Delta s_i - \frac{1}{2} \mathrm{Var}(\beta)\, \Delta \mathrm{Var}(U)_{\bar{\beta}}}
\end{equation}
Using this result to average over the ensemble of trajectories gives us the integrated fluctuation theorem
\begin{equation}
\left\langle e^{\Delta s_i - \frac{1}{2} \mathrm{Var}(\beta)\, \Delta \mathrm{Var}(U)_{\bar{\beta}}}\right\rangle_{P^{\bar{\beta}}} =1
\end{equation}
By using Jensen inequality, we finally obtain
\eq{
\langle \Delta \bar{s}_i \rangle_{\bar{\beta}} \geq \frac{1}{2} \mathrm{Var}(\beta)\, \Delta \mathrm{Var}(U)_{\bar{\beta}}
\label{eq:37}
}
{So depending on whether $\Delta \mathrm{Var}(U)_{\bar{\beta}}$ is positive or negative (i.e., whether the internal energy variance is increased or decreased during the experiment), the effective entropy production will be positive -- or negative.}


{ Finally, by plugging $\Delta s_i^{\bar{\beta}} = \bar{\beta}(w -\Delta F(\bar{\beta}))$ into \cref{eq:si}
(where $F(\bar{\beta})$ is the equilibrium free energy for temperature $\bar{\beta}$), we obtain
\begin{equation}
\Delta \bar{s}_i = \bar{\beta} \left( w - \Delta F(\bar{\beta})\right) + \frac{1}{2} \mathrm{Var}(\beta)\, \Delta \mathrm{Var}(U)_{\bar{\beta}}\, .
\end{equation}
By taking the ensemble average under the effective probability distribution, we obtain
\begin{equation}\label{eq:w}
\bar{W} \geq \Delta F(\bar{\beta}) - \frac{1}{2} \frac{\mathrm{Var}(\beta)}{\bar{\beta}} \Delta \mathrm{Var}(U)_{\bar{\beta}}\, .
\end{equation}
Next, we can expand the effective equilibrium free energy as
\begin{equation}
\bar{F} = F(\bar{\beta}) + \frac{1}{2} \mathrm{Var}(\beta) \frac{\partial^2 F}{\partial \beta^2}|_{\bar{\beta}}
\end{equation}
where
\begin{eqnarray}
\frac{\partial^2 F}{\partial \beta^2}|_{\bar{\beta}} &=& \frac{\partial^2}{\partial \beta^2}\left(-\frac{1}{\beta} \ln Z(\beta)\right)|_{\bar{\beta}} \nonumber\\
&=& \frac{2 F(\bar{\beta})}{\bar{\beta}^2} - \frac{2 U(\bar{\beta})}{\bar{\beta}^2} -\frac{\mathrm{Var}(U)_{\bar{\beta}}}{\bar{\beta}}
\label{eq:f}
\end{eqnarray}
By expressing $F(\bar{\beta})$  from Eq. \eqref{eq:f}
and plugging it into \cref{eq:w}, we obtain
\begin{equation}
\bar{W} \geq  \Delta \bar{F} + \bar{\beta} m_2(\beta)\,  \Delta S(\bar{\beta})
\end{equation}
}

\section{Optimal work extraction from a system with uncertain temperature}
\label{sec:first_minimal_EP_section}


In this and the following section we consider the foundational issue of the minimal thermodynamic
costs necessary to evolve an (uncertain) initial distribution to a (perfectly known) desired ending distribution, by
appropriate choice of a (perfectly known) time-dependent Hamiltonian.
For simplicity, we also assume
in both sections that we know with certainty that there is only
one reservoir during the system's evolution, and there are no particle interchanges between the system and that
reservoir during that evolution. (So the reservoir is a heat bath.)

While we can impose an arbitrary
time-dependent Hamiltonian $H_t(x)$ on the system, since we assume
that the system obeys LDB at all times with respect to that Hamiltonian, the rate matrix at any given moment
will depend on the temperature. For completeness, we assume that we have zero uncertainty about this dependence,
i.e., there is a single-valued function mapping $(H_t(x), \beta) \rightarrow K_{x,x'}(t)$ which
we can set arbitrarily (subject to the constraint of LDB), and which we know with zero uncertainty.
Accordingly, our choice of $H_t$ would determine the full function $K_{x,x'}(t)$ with zero uncertainty,
if it weren't for uncertainty about the temperature.

We start in this section by considering the case where in addition to being uncertain about the
initial distribution, we are also uncertain about the temperature of the heat reservoir coupled to the
system during its evolution. We further suppose that the (uncertain) initial distribution is
specified by the temperature of the heat reservoir that will govern the system's evolution
from that distribution. However, we allow this specification to
be flexible; we do not require that the initial distribution be an equilibrium for the heat reservoir.


Formally, in this section we suppose that
we have an initial set of distributions, one for each allowed temperature $\beta$, written as $p_{x}(t_i | \beta)$.
As mentioned above, we do \textit{not} assume that each of those distributions is at equilibrium
for the associated temperature.
Physically, we imagine a set of different physical apparatuses, each connected to its own reservoir with (inverse) temperature $\beta$,
and each with its own distribution, $p_{x}(t_i | \beta)$, where we randomly pick one of those
apparatuses, without knowing which one we have picked, by randomly sampling a distribution $\ud F_\beta = f(\beta) \ud \beta$.
Write the associated $\beta$-averaged initial distribution over $x$ as
\eq{
\bar{p}_{x}(t_i) = \sum_\beta f(\beta) p_{x}(t_i | \beta)
}

%
After choosing one of the apparatuses this way, we apply some protocol, ending with some effective
distribution
\eq{
\bar{p}_x(t_f) = \sum_\beta f(\beta) p_x(t_f |\beta)
}
We also have a \textbf{target} ending distribution, ${p^*}_x$, and want to evolve
$\bar{p}_x(t_i)$ to that target distribution. We are interested in minimizing the (dissipated) work required to do this.
To simplify the analysis, we assume that there is one and only one Hamiltonian $H^*_{t_f}(x)$ such that
the target distribution is a $\beta$-average of the associated equilibrium distribution for Hamiltonian $H^*_{t_f}$, i.e.,
%
\eq{
p^*_x &= \sum_\beta f(\beta) \dfrac{e^{-\beta H^*_{t_f}(x)}}{Z(H^*_{t_f}, \beta)}  \\
	&= \sum_{\beta} f(\beta) \dfrac{e^{-\beta H^*_{t_f}(x)}} {\sum_{x'} e^{-\beta H^*_{t_f}(x')}}
\label{eq:1}
}

If the protocol changes $p_x(t_i) \rightarrow p^*_x$ in finite time, then in general it will not end
with each distribution $p_x(t_f | \beta)$ equal to the associated Boltzmann distribution. (So if we were to watch the
system evolve after the ending time $t = 1$, we would see it change from $p^*$.) Even if the protocol
takes infinite time though, in general there will be more than one protocol that sends the initial distribution $p_x(t_i)$ to $p_x(t_f) = p^*_x$.
Whatever the protocol is, for each $\beta$ there will be an associated amount of EP that would arise if
that apparatus were the one chosen in our initial sampling process.
The $\beta$-average of those EPs is the $\beta$-average of the potentially
extractable work that will be dissipated due to our not being able to
tailor the protocol to match the actual apparatus and its inverse temperature $\beta$.
Our goal is to find which protocol generates minimal $\beta$-averaged EP while ending with the target distribution.

In this section we restrict attention to a protocol that comprises a conventional
 \{quench; equilibrate; semistatically-evolve\} process~\cite{parrondo2015thermodynamics,hasegawa2010generalization}.
Note that in the intermediate step, after the initial quench, we wait an arbitrarily long time, to allow the system to thermalize for the
new (post-quench) Hamiltonian, for whatever $\beta$ the system happens to have. So whatever semi-static evolution the system undergoes
after that step, it will generate no EP, i.e., all of the $\beta$-averaged EP is generated in the thermalization step.


Define $H^*_{t_i}(x)$ as the post-quench Hamiltonian that minimizes that  $\beta$-averaged EP.
To solve for $H^*_{t_i}(x)$, first note that for each $\beta$, the EP generated as $P_{t_i}(x | \beta)$ relaxes to
the associated equilibrium is given the Kullback Liebler divergence between two distributions over $X$:
\eq{
D\left( p_X({t_i} | \beta) \;\bigg|\bigg|\; \dfrac{e^{-\beta H^*_{t_i}(X)}}{Z(H^*_{t_i}, \beta)}\right)
\label{eq:2}
}
The $\beta$-averaged EP is the $\beta$-average of this KL divergence. Since the $\beta$-averaged
entropy of $p_x(t_i|\beta)$ is independent of $H^*_{t_i}(x)$, this means that $H^*_{t_i}(x)$ is the minimizer of
\eq{
\sum_{\beta} f(\beta) \left[ \sum_x p_x(t_i | \beta) \beta H^*_{t_i}(x) + \ln Z(H^*_{t_i}, \beta)\right]
\label{eq:3333}
}

Differentiating \cref{eq:3333} with respect to $H^*_{t_i}(x)$ for
each $x$ gives a set of coupled equations, one for each $x$:
\eq{
\sum_{\beta} f(\beta)\beta  \left[p_x(t_i | \beta) -  \dfrac{e^{-\beta H^*_{t_i}(x)}} {\sum_{x'} e^{-\beta H^*_{t_i}(x')}} \right] = 0
}
or if we define $g(x) := \sum_{\beta} f(\beta)\beta  p_x(t_i | \beta)$,
\eq{
g(x) = \sum_{\beta} f(\beta)\beta \dfrac{e^{-\beta H^*_{t_i}(x)}} {\sum_{x'} e^{-\beta H^*_{t_i}(x')}}
\label{eq:3}
}
(Compare to \cref{eq:1}.)

The effective reversible work expended during the semi-static evolution is the change in equilibrium (!)
free energy during the semi-static evolution,
\eq{
W^{ss} &:= \sum_\beta f(\beta) \left[F_\beta(H^*_{t_f}) - F_\beta(H^*_{t_i})\right] \\
	&= \sum_\beta f(\beta) \beta^{-1} \left[\ln Z(\beta, H^*_{t_i}) - \ln Z(\beta, H^*_{t_f})\right]
}
The associated change in effective energy is
\eq{
\Delta E^{ss} := \sum_{\beta, x} f(\beta) \left[H^*_{t_f}(x)  \dfrac{e^{-\beta H^*_{t_f}(x)}}{Z(H^*_{t_f}, \beta)} -
H^*_{t_i}(x)  \dfrac{e^{-\beta H^*_{t_i}(x)}}{Z(H^*_{t_i}, \beta)} \right]
}
The difference of those is the total heat during the semi-static evolution, $Q^{ss}$.

If we plug the optimal $H^*_{t_i}(x)$ solving \cref{eq:3} into \cref{eq:2}, we get a formula for the $\beta$-averaged dissipated work during the full protocol, $W^{diss}$.
In addition, if we write the initial, pre-quench Hamiltonian as $H^-_{t_i}(x)$, then the work during the quench is
\eq{
W^q := \sum_{\beta, x} f(\beta)p_x(t_i | \beta)\left[H^*_{t_i}(x) -  H^-_{t_i}(x)\right]
}
(In general, the dissipated work during the full protocol will differ from the actual work, since the change in entropy
as the system relaxes to its equilibrium distribution is work that could have been extracted but wasn't.)

If we add $W^{q}$ to $W^{ss}$,
we get the total reversible work during the full protocol. Now adding the $\beta$-averaged dissipated work
gives the total work during the full protocol,
\eq{
W^f = W^{ss} + W^q + W^{diss}
}
which provides a bound to the question motivating our analysis, of how much work can be extracted. (The
reason this is only a bound is that we have not proven that in the scenario considered here, the
 \{quench; equilibrate; semistatically-evolve\} process is optimal.)

In addition, the total change in internal energy during the full protocol is
\eq{
\Delta E^f &:= \Delta E^{ss} + W^q   \nonumber \\
	&\qquad + \sum_{\beta, x} f(\beta)H^*_{t_i}(x)\left[ \dfrac{e^{-\beta H^*_{t_i}(x)}}{Z(\beta, H^*_{t_i})} - p_x(t_i | \beta)\right] \\
	&=  W^q +
	 \sum_{\beta, x} f(\beta) \left[H^*_{t_f}(x)  \dfrac{e^{-\beta H^*_{t_f}(x)}}{Z(H^*_{t_f}, \beta)} - p_x(t_i | \beta) \right]
}
Now recall that in conventional, equilibrium thermodynamics, if we move a system isothermally and semi-statically from one equilibrium to another,
the difference between the work expended and the change in internal energy is the change in entropy (up to a multiplicative
factor given by the temperature). We would like an analogous interpretation to hold here. That means we
want to be able to interpret the difference $W^f -  \Delta E^f$ as the change in a generalized entropy during the full protocol, i.e., we want
there to be some functional $S^\dagger$ such that we can write
\eq{
& S^\dagger(p({t_i})) - S^\dagger(p^*) \nonumber \\
&\;\; = W^{diss} + W^{ss} -  \sum_{\beta, x} f(\beta) \left[H^*_{t_f}(x)  \dfrac{e^{-\beta H^*_{t_f}(x)}}{Z(H^*_{t_f}, \beta)} - p_x(t_i | \beta) \right]
\label{eq:57}
}

\section{Dynamics of the thermodynamic value of information}
\label{sec:second_minimal_EP_section}

We now investigate a more controlled version of the scenario considered in the previous
section. Just like in the previous section, here we are uncertain about the initial distribution. However,
in contrast to the scenario considered in the previous section,
here we are certain that the system evolves while coupled to a (single) heat reservoir with temperature $\beta = 1$.
To emphasize this distinction, here we index the initial set of distributions by an uncertain parameter $\alpha$ (not necessarily
equal to a temperature), writing it as $p_x(t_i | \alpha)$. So the $\alpha$-averaged distribution over $x$ is
\eq{
p_x(t_i) = \sum_\alpha f(\alpha) p_x(t_i | \alpha)
}

Physically, we imagine a process where we first randomly sample $f(\alpha)$, choosing one of the distributions, but don't
know which one was chosen. As in the previous section,
we want to apply a standard \{quench; equilibrate; semistatically-evolve\} process to that system,
in order to change the (unknown) distribution
$p_x(t_i | \alpha)$, whatever it is, to the (perfectly known) desired ending distribution $p_x(t_f) = P^*_x$ no matter what $\alpha$ is.
However, in contrast to the analysis of the previous section, here we consider the case where the evolution takes place
while the system is coupled to some single fixed heat reservoir at temperature $k_B T = 1$, no matter what $\alpha$ is.
In addition, we assume that there is a (perfectly known) delay between when the initial distribution is generated and when
the  \{quench; equilibrate; semistatically-evolve\} process begins, during which time the system evolves according
to a known rate matrix.

Our ultimate interest is in how the length of that delay affects the EP. To begin though, we consider
the simple case where the delay is zero, i.e., the  \{quench; equilibrate; semistatically-evolve\} process begins
immediately after the initial distribution is randomly generated.
First, note that if the distributions $p_x(t_i | \alpha)$ are not all the same, there is no single Hamiltonian $H(x)$ we can quench
to so that every $p_x(t_i | \alpha)$ (one for each $\alpha$) is at equilibrium for that $H(x)$. So
as in the previous section, whatever $H(x)$
we quench to will mean that the equilibration step, before the semistatic evolution,
will result in a nonzero $\alpha$-average of the (conventionally defined) EP of each of the
separate distributions $p_x(t_i | \alpha)$ as that distribution relaxes to the equilibrium distribution
defined by $H(x)$ and that $\alpha$.
%
So all the EP arises while the system is equilibrating.

Define  $H^\dagger(x)$ as the Hamiltonian that minimizes the $\alpha$-average of this initial EP.
To solve for $H^\dagger_{t_i}$, first note that for each $\alpha$, whatever Hamiltonian $H$ we use,
the EP generated as $p_x(t_i | \alpha)$ relaxes to equilibrium is
\eq{
D\left( p_x(t_i| \alpha) \;||\; \dfrac{e^{-H(x)}}{Z(H)}\right)
}
where as usual the partition function is
\eq{
Z(H) := \sum_x e^{-H(x)}
}

The $\alpha$-averaged EP is the $\alpha$-average of this KL divergence. Since the $\alpha$-averaged
entropy of $P_{t_i}(x|\alpha)$ is independent of $H(x)$, $H^\dagger_{t_i}(x)$ is the minimizer of
\eq{
\sum_x p_x({t_i}) \left[ H(x) + \ln Z(H)\right] = \ln Z(H) + \sum_x p_x({t_i})  H(x)
\label{eq:3333}
}
where
\eq{
p_x({t_i}) := \sum_\alpha f(\alpha) p_x(t_i | \alpha)
}

Differentiating \cref{eq:3333} with respect to $H(x)$ for
each $x$ gives
\eq{
p_x{t_i} &=  \dfrac{e^{- H(x)}} {\sum_{x'} e^{- H(x')}}
}
So
\eq{
H^\dagger_{t_i}(x) = -\ln P_x({t_i})
}
Note that this is the exact result one would have derived if one had simply marginalized over
the uncertainty about the initial distribution and then applied the conventional
calculation of what quenched Hamiltonian results in minimal EP.

%

Write $S\left(P_{t_i}(X | \alpha)\right)$ for the conditional entropy of $X$ given $\alpha$ under
joint distribution $P_{t_i}(X, \alpha) = f(\alpha) p_{t_i}(X| \alpha)$.
Then we can write the EP --- all generated in the stage where the system equilibrates --- as
\eq{
&\sum_\alpha P(\alpha) D\left( P_{t_i}(X | \alpha) \;||\; \dfrac{e^{-H^\dagger_{t_i}(X)}}{Z(H^\dagger_{t_i})}\right) \nonumber \\
		& \qquad\qquad \qquad= \langle H^\dagger_{t_i} \rangle_{P_{t_i}} - S\left(P_{t_i}(X | \alpha)\right) +  \ln Z(H^\dagger_{t_i}) \\
		& \qquad\qquad \qquad=  S\left(P_{t_i}(X)\right) -  S\left(P_{t_i}(X | \alpha)\right) \\
	    & \qquad\qquad \qquad= I_{P_{t_i}}(X ; \alpha)
}
which is just the Jensen Shannon divergence of the set of distributions $\{P(x | \alpha)\}$, weighted according to $P(\alpha)$.

This Jensen Shannon divergence is just the usual expression for the ``thermodynamic value of information'',
i.e., for the amount that the minimal extractable work changes if
we can measure some statistic $\alpha$ concerning the initial distribution of the state of the system compared to how much
we can extract if we can't measure that statistic.


%
%

As a variation of the scenario considered above, suppose that we are delayed in when we
can start the quench process, until some time $\tau > 0$ after $P(\alpha)$ is sampled.
In other words, suppose that after $\alpha$ is randomly chosen, but before the quench process starts,
the system evolves according to a known rate matrix, $W^{x'}_x(t)$, that is
the same for all $\alpha$, up to some time $\tau$, when the quench process starts.
If we take the time-derivative of this value of information with respect to $\tau$
we get
\begin{widetext}
\eq{
\dfrac{d  I_{P_\tau}(X ; \alpha)}{d\tau} &= \dfrac{d S\left(P_\tau(X)\right)}{d\tau} -  \dfrac{d S\left(P_\tau(X | \alpha)\right)}{d\tau} \\
	&=  \sum_{\alpha} P(\alpha) \sum_{x,x'} W^{x'}_x(\tau) P_\tau(x' | \alpha) \left[ \ln \dfrac{\sum_{\alpha'} P({\alpha'}) W^{x'}_x(\tau)
			P_\tau(x' | {\alpha'})}{\sum_{\alpha'} P({\alpha'}) W^{x}_{x'}(\tau) P_\tau(x | {\alpha'})}
					- \ln \dfrac{W^{x'}_x(\tau) P_\tau(x' | \alpha)}{W^{x}_{x'}(\tau) P_\tau(x | \alpha)} \right]
\label{eq:10}
}
\end{widetext}
\noindent
This difference in derivatives of entropies equals the difference between two EP rates,
distinguished by whether we know $\alpha$ or not; the EF rate
doesn't depend on our knowledge about $\alpha$, and so
the two EF rates (one for knowing $\alpha$, one for averaging over $\alpha$) cancel.

In addition, if we multiply and divide by $P(\alpha)$ inside the rightmost logarithm in \cref{eq:10},
and then apply the log-sum inequality, we see that $I_{P_\tau}(X ; \alpha)$ is monotonically increasing in time. (We can derive
the same result with the data-processing inequality.) Physically, this means that the minimal dissipated work that would
arise if we used our \{quench; equilibrate; semistatically-evolve\} process grows the longer we wait to use it,
assuming the system is undergoing $\alpha$-independent dynamics.

Next, suppose that $\alpha$ indexes not just the distribution, but also the rate matrix, i.e., suppose that we are unsure about the
dynamics
as well as the distribution. Then we must write $W^{x'}_{x}(\alpha, \tau)$
instead of $W^{x'}_x(\tau)$, and the dynamics of the value of information becomes
\begin{widetext}
\eq{
\dfrac{d  I_{P_\tau}(X ; \alpha)}{d\tau}
	&=  \sum_{\alpha} P(\alpha) \sum_{x,x'} W^{x'}_x(\alpha, \tau) P_\tau(x' | \alpha) \left[ \ln \dfrac{\sum_{\alpha'} P({\alpha'}) W^{x'}_x(\alpha', \tau) P_\tau(x' | {\alpha'})}{\sum_{\alpha'} P({\alpha'}) W^{x}_{x'}(\alpha', \tau) P_\tau(x | {\alpha'})}
					- \ln \dfrac{W^{x'}_x(\alpha, \tau) P_\tau(x' | \alpha)}{W^{x}_{x'}(\alpha, \tau) P_t(x | \alpha)} \right]
\label{eq:11}
}
\end{widetext}
Now the data-processing inequality doesn't apply. However, we can still use the log-sum inequality, to again
conclude that $I_{P_\tau}(X ; \alpha)$ is monotonically increasing in time.

A natural question is whether adding uncertainty to the rate matrix can cause the rate of increase in time of $I_{P_\tau}(X ; \alpha)$ to shrink, e.g.,
formalized as whether the following quantity can be negative:
\begin{widetext}
\eq{
\sum_{\alpha} P(\alpha) \sum_{x,x'} \left[W^{x'}_x(\alpha, \tau) P_\tau(x' | \alpha) \ln \dfrac{\sum_{\alpha'} P({\alpha'}) W^{x'}_x(\alpha', \tau) P_\tau(x' | {\alpha'}) }{\sum_{\alpha'} P({\alpha'}) W^{x}_{x'}(\alpha', \tau) P_\tau(x | {\alpha'})}
					-    \overline{W}^{x'}_x(\tau) P_\tau(x' | \alpha) \ln \dfrac{\sum_{\alpha'} P({\alpha'}) \overline{W}^{x'}_x(\tau)P_\tau(x' | {\alpha'}) } {\sum_{\alpha'} P({\alpha'}) \overline{W}^{x'}_x(\tau)  P_\tau(x | {\alpha'}) } \right]
\label{eq:12}
}
\end{widetext}
where $\overline{W}^{x'}_x(\tau) := \sum_{\alpha''} P(\alpha'')W^{x}_{x'}(\alpha'', \tau)$.
In the special case where all distributions $P(x | \alpha)$
are the same, so that the only uncertainty is in the rate matrix, the expression in \cref{eq:12} equals $0$.
Nonetheless, by applying the log-sum inequality to \cref{eq:11} for $P_t(x | \alpha)$ independent of $\alpha$ we see that the rate of increase of the value of information is still non-negative,
even when our only uncertainty concerns the rate matrix.

\section{Discussion and Future work}

In any real-world experimental test of a system there are three major types of uncertainty:
Uncertainty about the state of the system; uncertainty about the state of the external environment that the system is interacting with;
uncertainty about the parameters of the equations governing the dynamics of the system and its interaction
with its external environment. In (classical) stochastic thermodynamics, the first type of uncertainty is addressed
by replacing specification of the system's state, and the second one is typically addressed by assuming the environment
is infinite, at equilibrium, and evolving far faster than does the system (``separation of time scales''). In essence,
the entire field of stochastic thermodynamics concerns the consequences of those two types of uncertainty
for the dynamics of energy and particle counts in the system. However, very
little attention (if any) has been paid so far to the third type of uncertainty. Here we begin an investigation
of the consequences of that third type of uncertainty, showing that it entails major modifications to the
standard results previously derived in stochastic thermodynamics.

Our investigations have only scratched the surface of issues involved with this third type of uncertainty.
Some of the more immediate questions to be addressed in future work include:

%

\begin{enumerate}
\item For pedagogical simplicity we did not explicitly model uncertainty in the energy spectrum and / or associated
dynamics of the rate matrix in the analysis in \cref{sec:3}. Future work involves analyzing the relationship
between the thermodynamic consequences of those uncertainties and the thermodynamic consequences of the uncertainties that we
did explicitly consider in \cref{sec:3}.

\item How does the analysis in \cref{sec:4} change if we include low-order expansions of other
parameters besides temperature?

\item The approach outlined in \cref{sec:first_minimal_EP_section} defines generalized entropy using the total work expended during the
\{quench; equilibrate; semistatically-evolve\} process that takes $(P_{t_i}(x), H^-_{t_i}(x)) \rightarrow ( P^*(x), H^*_{t_f}(x))$
with minimal \textit{dissipated} work. Is that process the same as the one that minimizes the \textit{total} work during any (quench-thermalize-evolve) process that takes $(P_{t_i}(x), H^-_{t_i}(x)) \rightarrow ( P^*(x), H^*_{t_f}(x))$?

\item Is there any alternative protocol which, like the \{quench; equilibrate; semistatically-evolve\} process discussed above is ignorant of $\beta$, and which also maps $P_{t_i}(x) \rightarrow P^*(x)$ in infinite time,
but has less $\beta$-averaged EP than the quench-thermalize-evolve protocol?

\item Is there any such process which finishes in finite time, so that $P_{t_f}(x | \beta)$ is not an equilibrium
distribution for $H_{t_f}(x)$ for all $\beta$?

\item Is there any alternative protocol to the  \{quench; equilibrate; semistatically-evolve\} process considered in
\cref{sec:second_minimal_EP_section}
which is also ignorant of $\alpha$, and which also maps each possible initial distribution $P_{t_i}(x | \alpha)$ to $P_{t_f}(x)$,
but has less $\alpha$-averaged EP than $ I_{P_{t_i}}(X ; \alpha)$?

%
\end{enumerate}

%
%

\bibliographystyle{amsplain}
\bibliography{uncertain_thermo_refs.main,uncertain_thermo_refs_2,../../../../LANDAUER.Shared.2016/thermo_refs.main}

\newcommand{\arXiv}[2]{\href{http://arxiv.org/abs/#1}{arXiv:#1 #2}}
\providecommand{\bysame}{\leavevmode\hbox to3em{\hrulefill}\thinspace}
\providecommand{\MR}{\relax\ifhmode\unskip\space\fi MR }
\providecommand{\MRhref}[2]{%
  \href{http://www.ams.org/mathscinet-getitem?mr=#1}{#2}
}
\providecommand{\href}[2]{#2}
\begin{thebibliography}{10}

\bibitem{beck2003superstaitstic}
C.~Beck and E.G.D. Cohen, \emph{Superstatistics}, Physica A \textbf{322}
  (2003), 267--275.

\bibitem{bisker2017hierarchical}
Gili Bisker, Matteo Polettini, Todd~R Gingrich, and Jordan~M Horowitz,
  \emph{Hierarchical bounds on entropy production inferred from partial
  information}, Journal of Statistical Mechanics: Theory and Experiment
  \textbf{2017} (2017), no.~9, 093210.

\bibitem{davis2018temperature}
Sergio Davis and Gonzalo Guti{\'e}rrez, \emph{Temperature is not an observable
  in superstatistics}, Physica A: Statistical Mechanics and its Applications
  \textbf{505} (2018), 864--870.

\bibitem{esposito2010entropy}
Massimiliano Esposito, Katja Lindenberg, and Christian Van~den Broeck,
  \emph{Entropy production as correlation between system and reservoir}, New
  Journal of Physics \textbf{12} (2010), no.~1, 013013.

\bibitem{esposito2010three}
Massimiliano Esposito and Christian Van~den Broeck, \emph{Three faces of the
  second law. i. master equation formulation}, Physical Review E \textbf{82}
  (2010), no.~1, 011143.

\bibitem{fiore2019entropy}
Carlos~E. Fiore and M\'ario~J. de~Oliveira, \emph{Entropy production and heat
  capacity of systems under time-dependent oscillating temperature}, Phys. Rev.
  E \textbf{99} (2019), 052131.

\bibitem{garcia2011superstatistics}
Vladimir Garc{\'\i}a-Morales and Katharina Krischer, \emph{Superstatistics in
  nanoscale electrochemical systems}, Proceedings of the National Academy of
  Sciences \textbf{108} (2011), no.~49, 19535--19539.

\bibitem{hanel2011generalized}
Rudolf Hanel, Stefan Thurner, and Murray Gell-Mann, \emph{Generalized entropies
  and the transformation group of superstatistics}, Proceedings of the National
  Academy of Sciences \textbf{108} (2011), no.~16, 6390--6394.

\bibitem{hasegawa2010generalization}
H-H Hasegawa, J~Ishikawa, K~Takara, and DJ~Driebe, \emph{Generalization of the
  second law for a nonequilibrium initial state}, Physics Letters A
  \textbf{374} (2010), no.~8, 1001--1004.

\bibitem{kawai_dissipation:_2007}
R.~Kawai, J.~M.~R. Parrondo, and Christian Van~den Broeck, \emph{Dissipation:
  the phase-space perspective}, Physical Review Letters \textbf{98} (2007),
  no.~8, 080602.

\bibitem{matsuo2000stochastic}
Miki Matsuo and Shin ichi Sasa, \emph{Stochastic energetics of non-uniform
  temperature systems}, Physica A \textbf{276} (2000), no.~1, 188--200.

\bibitem{Note1}
We note though there is a substantial literature in which adopts an ``inclusive
  Hamiltonian'' framework, in which the external environment is finite, and the
  joint dynamics of the system-environment is explicitly modeled. This has been
  explored either with explicit Hamiltonian dynamics, with only randomness in
  the initial state~\cite
  {kawai_dissipation:_2007,esposito2010entropy,ptaszynski2018first} or at least
  approximate Hamiltonian dynamics~\cite
  {seifert2016first,strasberg2017stochastic,talkner2020HMF,talkner2020comment,strasberg2020measurability}.

\bibitem{Note2}
We note though that it is already known that if we do not account for all
  thermodynamic reservoirs, we invariably undercount the total of entropy
  production in a process~\cite {esposito2010three}.

\bibitem{Note3}
{\protect \color {black}Uncertainty in the energy spectrum and / or the
  time-dependence of the rate matrix can also be included in the analysis, at
  the cost of more complex notation.}

\bibitem{parrondo2015thermodynamics}
Juan~MR Parrondo, Jordan~M Horowitz, and Takahiro Sagawa, \emph{Thermodynamics
  of information}, Nature Physics \textbf{11} (2015), no.~2, 131--139.

\bibitem{ptaszynski2018first}
Krzysztof Ptaszy{\'n}ski, \emph{First-passage times in renewal and nonrenewal
  systems}, Physical Review E \textbf{97} (2018), no.~1, 012127.

\bibitem{sattin2006bayesian}
Fabio Sattin, \emph{Bayesian approach to superstatistics}, The European
  Physical Journal B-Condensed Matter and Complex Systems \textbf{49} (2006),
  no.~2, 219--224.

\bibitem{seifert2012stochastic}
Udo Seifert, \emph{Stochastic thermodynamics, fluctuation theorems and
  molecular machines}, Reports on Progress in Physics \textbf{75} (2012),
  no.~12, 126001.

\bibitem{seifert2016first}
\bysame, \emph{First and second law of thermodynamics at strong coupling},
  Physical Review Letters \textbf{116} (2016), no.~2, 020601.

\bibitem{shiraishi_ito_sagawa_thermo_of_time_separation.2015}
Ito S. Kawaguchi~K. Shiraishi, N. and T.~Sagawa, \emph{Role of
  measurement-feedback separation in autonomous {M}axwell's demons}, New
  Journal of Physics (2015).

\bibitem{strasberg2017stochastic}
Philipp Strasberg and Massimiliano Esposito, \emph{Stochastic thermodynamics in
  the strong coupling regime: An unambiguous approach based on coarse
  graining}, Physical Review E \textbf{95} (2017), no.~6, 062101.

\bibitem{strasberg2019non}
\bysame, \emph{Non-markovianity and negative entropy production rates},
  Physical Review E \textbf{99} (2019), no.~1, 012120.

\bibitem{strasberg2020measurability}
\bysame, \emph{Measurability of nonequilibrium thermodynamics in terms of the
  hamiltonian of mean force}, Physical Review E \textbf{101} (2020), no.~5,
  050101.

\bibitem{talkner2020comment}
Peter Talkner and Peter H{\"a}nggi, \emph{Comment on" measurability of
  nonequilibrium thermodynamics in terms of the hamiltonian of mean force''},
  arXiv preprint arXiv:2006.16592 (2020).

\bibitem{van2015ensemble}
Christian Van~den Broeck and Massimiliano Esposito, \emph{Ensemble and
  trajectory thermodynamics: A brief introduction}, Physica A: Statistical
  Mechanics and its Applications \textbf{418} (2015), 6--16.

\bibitem{wolpert1993use}
David~H Wolpert et~al., \emph{On the use of evidence in neural networks},
  Advances in neural information processing systems (1993), 539--539.

\end{thebibliography}

\clearpage
\appendix
\section{Brief review of stochastic thermodynamics}

\begin{equation}
\dot{p}_x(t) = \sum_{x'} K_{x,x'} p_{x'}(t) = \sum_{x'} J_{x,x'}\, .
\end{equation}
Transition rates can be decomposed as
\begin{equation}
K_{x,x'} = \sum_{\nu=1}^{N} K_{x,x'}^{\nu}\, .
\end{equation}
The local equilibrium distribution is defined as
\begin{equation}\label{eq:eq}
\pi^{\nu}_x = \frac{\exp(-\beta_\nu (\epsilon_x - \mu_\nu n_x))}{Z^{\nu}}\, .
\end{equation}
Transition rates fulfill local detailed balance, so
\begin{equation}\label{eq:ldb}
\frac{K^{\nu} _{x,x'}}{K^{\nu}_{x'x}} = \frac{\pi^{\nu}_x}{\pi^{\nu}_{x'}}  = e^{-\beta_\nu[(\epsilon_x - \epsilon_{x'}) -\mu_\nu(n_x-n_{x'})]}\, .
\end{equation}
Thus, local flows
\begin{equation}
J_{x,x'}^{\nu} = K_{x,x'}^{\nu} p_{x'}(t) - K_{x'x}^{\nu} p_{x}(t)
\end{equation}
vanish for local equilibrium distributions.

We define internal energy of the system as
\begin{equation}
U = \sum_x p_x \epsilon_x.\end{equation}
The first law of thermodynamics can be formulated as follows:
\begin{equation}
\dot{U} = \dot{W} + \dot{Q} + \dot{W}_{chem}
\end{equation}
where
\begin{eqnarray}
\dot{W} &=& \sum_x p_x \dot{\epsilon}_x\\
\dot{Q} &=& \sum_{\nu} \dot{Q}^{\nu} = \sum_\nu  \sum_x \dot{p}_x \left(\epsilon_x-\mu_\nu n_x\right) \\
\dot{W}_{chem} &=& \sum_{\nu} \dot{W}_{chem}^{\nu} = \sum_\nu \sum_x \dot{p}_x  \mu_\nu n_x
\end{eqnarray}

Entropy of the system is defined as
\begin{equation}
S = - \sum_x p_x \ln p_x
\end{equation}
Second law of thermodynamics can be expressed as follows:
\begin{equation}\label{eq:st}
\dot{S} = \dot{S}_i + \dot{S}_e
\end{equation}
where
\begin{eqnarray}
\dot{S}_i &=& \frac{1}{2} \sum_{x,x'\nu} J_{x,x'}^{\nu} \ln \frac{w_{x,x'}^{\nu} p_x}{K_{x'x}^{\nu} p_{x'}} \geq 0\\
\dot{S}_e &=& \frac{1}{2} \sum_{x,x'\nu} J_{x,x'}^{\nu} \ln \frac{w_{x,x'}^{\nu}}{K_{x'x}^{\nu}} \nonumber\\
&=& \sum_\nu  \beta_\nu \dot{Q}^{\nu}\, .
\end{eqnarray}
\paragraph{Trajectory thermodynamics and fluctuation theorems:}
We consider a stochastic trajectory $x(\tau)$. We define stochastic energy as $\epsilon(\tau) = \epsilon_{x(\tau)}$. We also consider that particle levels $n_x$ are independent of time.  The first law of thermodynamics can be expressed on the trajectory level as
\begin{equation}
\frac{\ud}{\ud t} e(\tau) = \dot{w}(\tau) + \dot{q}(\tau) + \dot{w}_{chem}(\tau)
\end{equation}
where
\begin{eqnarray}
\dot{w} &=& \sum_x \delta_{x,x(\tau)} \dot{\epsilon}_{x(\tau)}\\
\dot{q} + \dot{w}_{chem}(\tau) &=& \sum_x \dot{\delta}_{x,x(\tau)} \epsilon_{x(\tau)}
\end{eqnarray}
It is important to stress that the first law of thermodynamics is the pure energetic balance which does depends neither on the number of reservoirs nor on their parameters. Heat and chemical work can be decomposed into contributions from particular reservoirs
\begin{eqnarray}
\dot{q} = \sum_\nu \dot{q}^\nu(\tau)= \sum_{\nu}  \dot{\delta}_{x,x(\tau)} (\epsilon_{x(\tau)}-\mu_x n_{x(\tau)})\\
 \dot{w}_{chem} = \sum_{\nu} \dot{w}^\nu_{chem}(\tau) = \sum_\nu  \dot{\delta}_{x,x(\tau)} \mu_x n_{x(\tau)}
\end{eqnarray}

Stochastic entropy along a trajectory $x(\tau)$ can be defined as
\begin{equation}
s_{x(\tau)} = - \ln p_{x(\tau)}\, .
\end{equation}
It is straightforward to show that stochastic entropy fulfills the second law of thermodynamics, i.e.,
\begin{equation}
\dot{s} = \dot{s}_i +  \dot{s}_e
\end{equation}
where
\begin{equation}
\dot{s}_e = \sum_\nu  \beta_\nu \dot{q}^\nu
\end{equation}
By averaging over the trajectory ensemble, we obtain the ensemble second law of thermodynamics in the form $\langle \dot{s}_i \rangle = \dot{S}_i \geq 0$.

Stochastic entropy production satisfies the well-known fluctuation theorem, which is a consequence of the following expression:
\begin{equation}\label{eq:dft}
\ln \frac{\mathcal{P}_{x(\tau)}}{\tilde{\mathcal{P}}_{\tilde{x}(\tau)}} = \Delta s_i
\end{equation}
where $\mathcal{P}_{x(\tau)}$ is the probability of stochastic trajectory $x(\tau)$ for a protocol $\lambda(\tau)$ and $\tilde{\mathcal{P}}_{\tilde{x}(\tau)}$ is the probability of observing time-reversed trajectory $\tilde{x}(\tau) = x(T-\tau)$ for a time-reversed protocol $\tilde{\lambda}(\tau) = \lambda(T-\tau)$. It is straightforward to show that $\Delta s_i$ obeys the detailed fluctuation theorem (DFT)
\begin{equation}
\frac{P(\Delta s_i)}{\tilde{P}(-\Delta s_i)} = e^{\Delta s_i}
\end{equation}
as well as integrated fluctuation theorem (IFT)
\begin{equation}
\left\langle e^{-\Delta s_i} \right\rangle = 1\, .
\end{equation}

\section{General form of transition matrix satisfying local detailed balance}

We now focus on the most general form of transition matrix which fulfills local detailed balance. We denote transition rate matrix as $K = \{K_{x,x'}\}_{x,x'}$ and probability distribution $p(t) = \{p_x(t)\}_x$. We start with the most general form of a transition matrix satisfying detailed balance, which is a special case of local detailed balance when a system is coupled to one reservoir. We define the equilibrium distribution $\pi =\{\pi_x\}_x$ as the solution of
\begin{equation}
K \pi = 0.
\end{equation}

We define $\Pi = \mathrm{diag}(\pi)$ where $\mathrm{diag}$ denotes a diagonal matrix with diagonal elements equal to elements of the vector. Requirement of detailed balance from Eq. \eqref{eq:ldb} can be in the matrix notation rewritten as
\begin{equation}\label{eq:ldbm}
K \Pi = \left(K \Pi\right)^T
\end{equation}

We consider the following decomposition of the transition rate matrix:
\begin{equation}\label{eq:sdm}
K = R \, \Pi^{-1}
\end{equation}
Then, we obtain from Eq. \eqref{eq:ldbm} that $R = R^T$, i.e., the matrix is symmetric. It is actually possible to show that Eq. \eqref{eq:sdm} is the most general decomposition of the transition rate matrix satisfying detailed balance. The normalization of the transition rate matrix is obtained in the diagonal elements of $R$. For a general matrix $R'$, the general transition rate matrix can be expressed as
\begin{equation}\label{eq:sdm}
K = R' \Pi^{-1} - \mathrm{diag} \left(R' \Pi^{-1} \cdot \textbf{1}\right)
\end{equation}
where $\cdot$ denotes standard application of matrix to a vector and $\textbf{1} = (1,\dots,1)$. In the following, we always assume that the matrix $R$ is chosen so that $K$ is normalized ($\sum_x K_{x,x'} = 0$). Consequently, the probability flow can be expressed as follows:
\begin{equation}
J_{x,x'} = R_{x,x'} \left(\frac{p_{x'}(t)}{\pi_{x'}}-\frac{p_x(t)}{\pi_x}\right)
\end{equation}
from which we immediately see that the system satisfies detailed balance.
A matrix satisfying local detailed balance (LDB-obeying matrix) is then a sum of matrices satisfying detailed balance, so
\begin{equation}
K = \sum_\nu K^\nu = \sum_{\nu=1}^N R^\nu \, \left(\Pi^\nu\right)^{-1}
\end{equation}

We now assume the local equilibrium distribution in the form of Boltzmann distribution. Then, $\Pi^\nu$ is parameterized by $\alpha^\nu= (\beta^\nu,\mu^\nu)$ (we consider that the energy levels are known and fixed). We also denote $\mathbf{R} = \{ R^\nu\}_{\nu=1}^N$, then a general transition matrix satisfying local detailed balance can be characterized by the following variables
\begin{equation}
K(N;\bt,\mathbf{R}) = \sum_{\nu=1}^N  R^\nu (\Pi^\nu)^{-1} - \mathrm{diag} \left(R^\nu (\Pi^\nu)^{-1} \cdot \textbf{1}\right)
\end{equation}

\end{document}